# 180°-twisted bilayer ReSe$_2$ as an artificial noncentrosymmetric semiconductor

S. Akatsuka[1], M. Sakano[1], T. Yamamoto[1], T. Nomoto[2], R. Arita[2,3], R. Murata[4], T. Sasagawa[4], K. Watanabe[5], T. Taniguchi[6], M. Kitamura[7,8], K. Horiba[8], K. Sugawara[9,10,11], S. Souma[10], T. Sato[9,10], H. Kumigashira[12], K. Shinokita[13], H. Wang,[13] K. Matsuda[13], S. Masubuchi[14], T. Machida[14], and K. Ishizaka [1,3]

[1]Quantum-Phase Electronics Center and Department of Applied Physics, The University of Tokyo, Bunkyo-ku, Tokyo, 113-8656, Japan
[2]Research Center for Advanced Science and Technology, The University of Tokyo, Meguro-ku, Tokyo 153-8904, Japan
[3]RIKEN Center for Emergent Matter Science (CEMS), Wako, Saitama, 351-0198, Japan
[4]Materials and Structures Laboratory, Tokyo Institute of Technology, Yokohama, Kanagawa, 226-8503, Japan.
[5]Research Center for Electronic and Optical Materials, National Institute for Materials Science, 1-1 Namiki, Tsukuba 305-0044, Japan
[6]Research Center for Materials Nanoarchitectonics, National Institute for Materials Science, 1-1 Namiki, Tsukuba 305-0044, Japan
[7]Photon Factory, Institute of Materials Structure Science, High energy Accelerator Research Organization (KEK), Tsukuba 305-0801, Japan
[8]Institute for Advanced Synchrotron Light Source, National Institute for Quantum Science and Technology (QST), Sendai 980-8579, Japan
[9]Department of Physics, Graduate School of Science, Tohoku University, Sendai 980-8578, Japan
[10]Advanced Institute for Materials Research (WPI-AIMR), Tohoku University, Sendai 980-8577, Japan
[11]Precursory Research for Embryonic Science and Technology (PRESTO), Japan Science and Technology Agency (JST), Tokyo 102-0076, Japan
[12]Institute of Multidisciplinary Research for Advanced Materials (IMRAM), Tohoku University, Sendai 980-8577, Japan
[13]Institute of Advanced Energy, Kyoto University, Uji, Kyoto 611-0011, Japan
[14]Institute of Industrial Science, The University of Tokyo, Meguro-ku, Tokyo 153-8505, Japan

**We have fabricated a 180°-twisted bilayer ReSe$_2$ by stacking two centrosymmetric monolayer ReSe$_2$ flakes in opposite directions, which is expected to lose spatial inversion symmetry. By the second harmonic generation and angle-resolved photoemission spectroscopy, we successfully observed spatial inversion symmetry breaking and emergent band dispersions. The band calculation shows the finite lifting of spin degeneracy (~50 meV) distinct from natural monolayer and bilayer ReSe$_2$. Our results demonstrate that the spin-momentum locked state, which leads to spintronic functions and Berry-curvature-related phenomena, can be realized even with the stacking of centrosymmetric monolayers.**

Advances in the fabrication techniques for exfoliated two-dimensional flakes and their van der Waals heterostructures have provided a platform for manipulating material symmetries through stacking order [1–4]. Among the various symmetries, spatial inversion symmetry is important in determining electronic structure and physical properties. One of the most representative examples is group VI transition metal dichalcogenide (TMD) semiconductors such as $MoS_2$. Monolayer $MoS_2$ has a non-centrosymmetric crystal structure with three-fold rotational symmetry, which lifts the spin degeneracy at the Brillouin zone corners by spin-orbit interactions [5–8], leading to spintronic functions and Berry-curvature-related phenomena [5–12]. The bilayer system has two distinct stacking order: 2H- and 3R-type, in which the adjacent layers are stacked with 180° and 0° twisting [13]. The former recovers the spatial inversion symmetry, and the net spin polarization cancels out [13,14]. The latter maintains the breaking of spatial inversion symmetry [13], and recently, emergent ferroelectricities in 0°-stacked bilayer systems have been reported [3,4].

On the other hand, there are examples where the stacking order, regardless of the stacking of centrosymmetric monolayers, breaks the spatial inversion symmetry in a bilayer system. $T_d$-$WTe_2$ [15–22] is a well-known striking material. The crystal structure of the monolayer $WTe_2$ is centrosymmetric and is classified as a distorted 1T-type ($CdI_2$-type) in the TMD family, which is defined by a network of edge-sharing $WTe_6$ octahedra [Fig. 1(a)]. The distinctive feature of the crystal symmetry of bulk $T_d$-$WTe_2$ is that adjacent layers stack in opposite directions in natural, in contrast to the usual 1T-type bulk TMD families (e.g. $HfSe_2$ [23–25], $TiSe_2$ [23,24,26], and $ReSe_2$ [27]) which have spatial inversion symmetry independent of the number of layers. This 180°-twisted stacking breaks the spatial inversion symmetry in bilayer $WTe_2$, where the spin-splitting of approximately 0.1 eV has been observed by micro-focused angle-resolved photoemission spectroscopy (μ-ARPES) [19,22]. Inspired by these previous studies, we can expect that, in principle, the magnitude of spin splitting could be controlled by artificially changing the stacking angle, even when stacking with centrosymmetric 1T-type TMD monolayers.

In this study, we focus on the layered semiconductor $ReSe_2$. Reflecting the strong spin-orbit interaction of the Re $5d$ orbitals, the effect of inversion symmetry breaking is expected to appear in the nonlinear optical properties and electronic band dispersions. The direct and indirect band gaps of bulk $ReSe_2$ are 1.40 eV [28] and 1.18–1.19 eV [29,30], respectively. $ReSe_2$ is known to have a weak interlayer coupling, making it possible to fabricate down to a monolayer by the mechanical exfoliation [31–34] with an exitonic direct band gap of 1.50 eV [32]. The overall electronic structure does not change significantly regardless of the number of layers, indicating weak van der Waals interactions between the layers [32,34,35]. $ReSe_2$ has a distorted 1T-type triclinic crystal structure (space group $P\bar{1}$), as shown in Figs. 1(a)–1(c). The opposed Se-triangle networks (shown in yellow and orange) sandwich the distorted Re layer characterized by Re zigzag chains (indicated by the light blue lines). As shown in Fig. 1(d), we schematically illustrate the concept of this study for the artificial fabrication of noncentrosymmetric bilayer $ReSe_2$. The essence extracted from the crystal structure of monolayer (1L) $ReSe_2$ is represented by two opposing Se-triangular networks (yellow and orange triangles) and the inversion center (black circle). In the natural bilayer case, an inversion center appeared between the layers. However, in the 180°-twisted 2L ReSe$_2$, the triangles face the same direction across the interlayer, such that the inversion center does not appear anywhere between the layers, shown as a dotted circle. Although there are translational degrees of freedom upon stacking in the *x* and *y* directions, the fabricated structure cannot have an inversion center in any cases. In 180°-twisted 2L ReSe$_2$, spin degeneracy is expected to be lifted in reciprocal space owing to spin-orbit interaction, which has the potential to give rise to spintronic functionalities and physical properties related to the Berry curvature.

The 180°-twisted 2L ReSe$_2$ sample was fabricated using an all-dry pick-up [36,37], tear-and-stack [38], and flip method [39] using an Elvacite2552C copolymer inside a glovebox chamber [40]. In addition, we also prepared the monolayer and natural bilayer ReSe$_2$ samples for control experiments. During fabrication, hexagonal boron nitride (hBN), graphite, and ReSe$_2$ flakes were sequentially picked using a polymer stamp. The assembled heterostructure was transferred to another polymer stamp at room temperature to turn it over. Then, it was dropped onto a SiO$_2$/Si substrate with a prepatterned metal electrode, as schematically shown in the inset of Fig. 2(a). Optical microscope images for 1L, 2L, and 180°-twisted 2L ReSe$_2$ are shown in Figs. 2(a)–2(c), respectively. The ReSe$_2$ flakes were approximately 10 μm in size, outlined with orange and green broken lines. In this study, we performed optical second harmonic generation (SHG) measurements at room temperature [40] to examine the breaking of the spatial inversion symmetry. In addition, to observe the emergent electronic band dispersions in 180°-twisted 2L ReSe$_2$ we performed μ-ARPES measurements using a photon energy of 100 eV and a spot size of 12 × 15 μm$^2$ at BL28 in the Photon Factory, KEK [41]. The total energy resolution was set at 35 meV. During the measurement, a sample manipulator temperature was kept below 20 K. The Fermi levels ($E_F$) were determined using polycrystalline gold electrically connected to the respective samples. The band structure calculations [40] were performed using the Vienna Ab initio Simulation Package (VASP) [42,43]. The crystal structures were determined by structure optimization taking into account the van der Waals correction with the DFT-D3 method [44]. The second-order susceptibility calculations were performed based on the tight-binding model constructed from the DFT electronic structures through Wannier90 code [45].

Figures 2(d)–2(f) show the polar-angle dependences of the normalized SHG signals from the 1L, 2L, and 180°-twisted 2L ReSe$_2$ samples. The orange circles indicate the measurement areas on the topmost ReSe$_2$ flakes (depicted with the orange frames). The SHG signals are detected only from 180°-twisted 2L; however, they are negligible from 1L and 2L reflecting the spatial inversion symmetry of natural ReSe$_2$ crystal independent of the number of layers. These experimental results are consistent with the description using simple structure models shown in Fig. 1(d). To compare with the experimental observation, we calculated second-order susceptibility $\chi_{xx}$ and $\chi_{yy}$, which represent the SHG response parallel to the incident light polarization [along *x* and *y*, respectively, as shown in Fig. 1(c)] [40]. The energy dependences of $\chi_{xx}$ and $\chi_{yy}$ for 1L, 2L, and 180°-twisted 2L are shown in Fig. 2(g). In contrast to the negligible $\chi_{xx}$ and $\chi_{yy}$ for 1L and 2L, the finite $\chi_{xx}$ and $\chi_{yy}$ were demonstrated for 180°-twisted 2L, indicating that the observed SHG signal was consistent with the design of the breaking of spatial inversion symmetry. The crystal structure of 180°-twisted 2L used for the calculations is shown in the supplementary material [40].

We performed ARPES measurements on 1L, 2L, and 180°-twisted 2L $ReSe_2$ to observe their two-dimensional electronic structures. Figures 3(a)–3(c) show the ARPES intensity mapping at the constant energies of $E - E_F = -1.3$ eV. The constant energy contours for the highest valence band (HVB) observed in Figs. 3(a)–3(c) clearly show anisotropic contours of the isoenergetic surface that are not closed along the $k_y$ direction, reflecting the quasi-one-dimensional Re-zigzag chain along $x$ as shown in Fig. 1(c). Figs. 3(d)–3(f) show the ARPES images along the $k_y$ (Γ-$M$, left side) and $k_x$ (Γ-$K$-$M$, right side) directions [Fig. 1(e)] for 1L, 2L, and 180°-twisted 2L $ReSe_2$, respectively. The ARPES intensities along the $k_x$ ($k_y$) direction shown in Figs. 3(d)–3(f) are symmetrized with respect to $k_x$ ($k_y$) = 0 for better visibility. In the energy region from the Fermi energy to $E = E_F - 0.8$ eV, which is omitted in Fig. 3, no bands originating from $ReSe_2$ are observed, whereas there are certain signals from graphite. This is consistent with previous studies of 1L and 2L $ReSe_2$ [33,34]. For better visualization of the band dispersions, curvature plots [46] for the respective ARPES images are shown in Fig. 3(g)–3(i). Comparing 1L, 2L, and 180°-twisted 2L, a clear difference became apparent when focusing on the HVB. A recent precise ARPES study revealed that the valence band maximum (VBM) of bulk $ReSe_2$ located off the high symmetry point along the $k_y$ direction ($k_y \sim 0.15$ Å$^{-1}$) at $k_z = \pi/c$ with upward convex-shaped dispersions [47,48]. As discussed below, the ARPES results demonstrate that the two-dimensional confinement of the electronic structure causes a shift in the position of the VBM. The detailed analysis of our result on 1L $ReSe_2$ shows that the HVB is almost flat with 40 meV dispersion within the $k = \pm 0.15$ Å$^{-1}$ range around the Γ-point, as indicated by the markers in Fig. 3(d) that represent the positions of intensity peaks. In contrast to 1L $ReSe_2$, 2L and 180°-twisted 2L $ReSe_2$ exhibited upward convex-shaped HVBs with the VBM at Γ-point as shown in Figs 3(e) and 3(f). Figures 3(j)–3(l) show the band dispersions obtained from first-principles band calculations with optimizing the crystal structure [40] for 1L, 2L, and 180°-twisted 2L $ReSe_2$, respectively. Compared to the calculation results using the fixed atomic coordinates of bulk crystal [27, 40], the calculation with structural optimization tends to form flatter HVBs around the Γ-point, which reproduces the observed band dispersions for 1L and 2L $ReSe_2$. Although the precise crystal structure of the 180°-twisted 2L $ReSe_2$ remains undetermined, and the calculation result assuming one of the possible crystal structures [40] appears plausible, as it also reproduces the ARPES results well.

Subsequently, we discuss the effect of 180°-twisted stacking on band dispersions in $ReSe_2$. In Figs 3(j)–3(l), we can observe that the number of bands in the band calculations doubles from 1L to 2L, and from 2L to 180°-twisted 2L $ReSe_2$, respectively. The former was owing to the bilayer splitting caused by doubling the number of atoms in a unit cell. However, the latter doubling was owing to the breaking of spatial inversion symmetry at 180°-twisted 2L induced by staggered stacking [Fig. 1(d)], leading to the lifting of spin degeneracy. The maximum spin-splitting energy was estimated to be approximately 70 meV in this calculation. To evaluate the possible spin-split band dispersions appearing in the experimental results, we reviewed the ARPES images of the 180°-twisted 2L ReSe$_2$ in Figs. 3(f) and 3(i). Note that the ARPES intensities from the non-overlapped and/or non-hybridized 1L ReSe$_2$ flakes were inevitable because the beam-spot size was comparable to the sample size. In Fig. 3(f), we showed guide for the eyes depicted with broken white curves representing the band dispersions of 1L ReSe$_2$ extracted from the ARPES image in Fig. 3(d). In addition to the HVB, a new band dispersion appeared in 180°-twisted 2L ReSe$_2$ as indicated by a white arrow in Fig. 3(i) is observed around $E - E_F = -1.9$ eV at the Γ-point where the band dispersion does not exist in the 1L ReSe$_2$ [Fig. 3(f)]. Comparison of the calculated band dispersions between 2L and 180°-twisted 2L ReSe$_2$ in Figs. 3(k) and 3(l), the corresponding band dispersions are well isolated from the other band dispersions. As expected from the symmetry requirement shown in Fig. 1(d), only 180°-twisted 2L ReSe$_2$ [Fig. 3(l)] exhibited spin-split band dispersion.

Finally, we focus on the experimental results of the band dispersion at 180°-twisted 2L ReSe$_2$ to examine whether there is a footprint of spin-split band dispersion. Magnified view of the calculated band dispersions of 2L and 180°-twisted 2L ReSe$_2$, corresponding to the area indicated by the black rectangles in Figs. 3(k) and 3(l), are shown in Figs. 4(a) and 4(b), respectively. We observed band splitting with lifting of the spin degeneracy only for 180°-twisted 2L ReSe$_2$. Spin-splitting is larger in the $k_y$ direction, with a maximum of approximately 50 meV. We performed calculations for three different $x$-, $y$- shifts on stacking and identified that the energies of spin splitting were almost similar [40]. The energy distribution curves (EDCs) extracted from the ARPES images at 2L and 180°-twisted 2L ReSe$_2$ in Figs. 3(e) and 3(f) are shown in Figs. 4(c) and (d), respectively. The corresponding energy cuts from $k = -0.12$ to $0.12$ Å$^{-1}$ are indicated by the red lines in Figs. 3(e) and 3(f). The EDCs of 180-deg.-twisted 2L ReSe$_2$, when comparing the EDC spectra at the Γ-point and away from the Γ-point, it appears that the latter exhibited a broader shape and indicative of two-peak structures. This observation is indicated by black circles representing the peak positions of the EDCs and shaded gray lines as visual guides. In contrast, the EDCs of 2L ReSe$_2$ also exhibited broad two-peak like structures but with parallel dispersions, as indicated by the black solid and open circle markers. The weak peaks indicated by open circles do not appear in the calculation and its origin is not clear. But considering rather broad and hazy ARPES image obtained from 2L-ReSe$_2$, it may be attributed to ARPES intensities originating from inhomogeneous areas *e.g.* with different chemical potentials. Considering the mechanism of the band splitting, direct observation of spin polarization on respective band dispersions detected by spin-resolved ARPES is a future challenge for this study.

In summary, we fabricated a 180°-twisted 2L ReSe$_2$ with the breaking of spatial inversion symmetry, whereas 1L and natural 2L ReSe$_2$ possessed symmetry. We have detected the SHG signal only in the 180°-twisted 2L ReSe$_2$, in a good agreement with the calculation, indicating the breaking of the spatial inversion symmetry. Our ARPES study also indicated the stacking-dependent emergent electronic band dispersions in these ReSe$_2$ thin flakes. Our study successfully demonstrated that it is possible to artificially induce peculiar physical properties such as ferroelectricity and Berry-curvature-related phenomena even by stacking inversion-symmetric two-dimensional crystals.

**ACKOWLDGEMENTS**

This research was partly supported by a CREST project (Grant No. JPMJCR18T1 and No. JPMJCR20B4) from the Japan Science and Technology Agency (JST), Japan Society for the Promotion of Science KAKENHI (Grants-in-Aid for Scientific Research) (Grants No. JP20H01834, No. JP20H05664, No. JP21H01012, JP21H01757, JP21H04652, No. JP21H05232, No. JP21H05233, No. JP21H05234, No. JP21H05235, No. JP21H05236, JP21K18181, No. JP22K18986, No. JP23H02052 and No. JP23H05469), JST SPRING (Grant No. JPMJSP2106), and JST PRESTO (Grant No. JPMJPR20A8). K.W. and T.T. acknowledge support from World Premier International Research Center Initiative (WPI), MEXT, Japan. This work was partly performed under the approval of the photon Factory Program Advisory Committee (Proposal No. 2023G088).

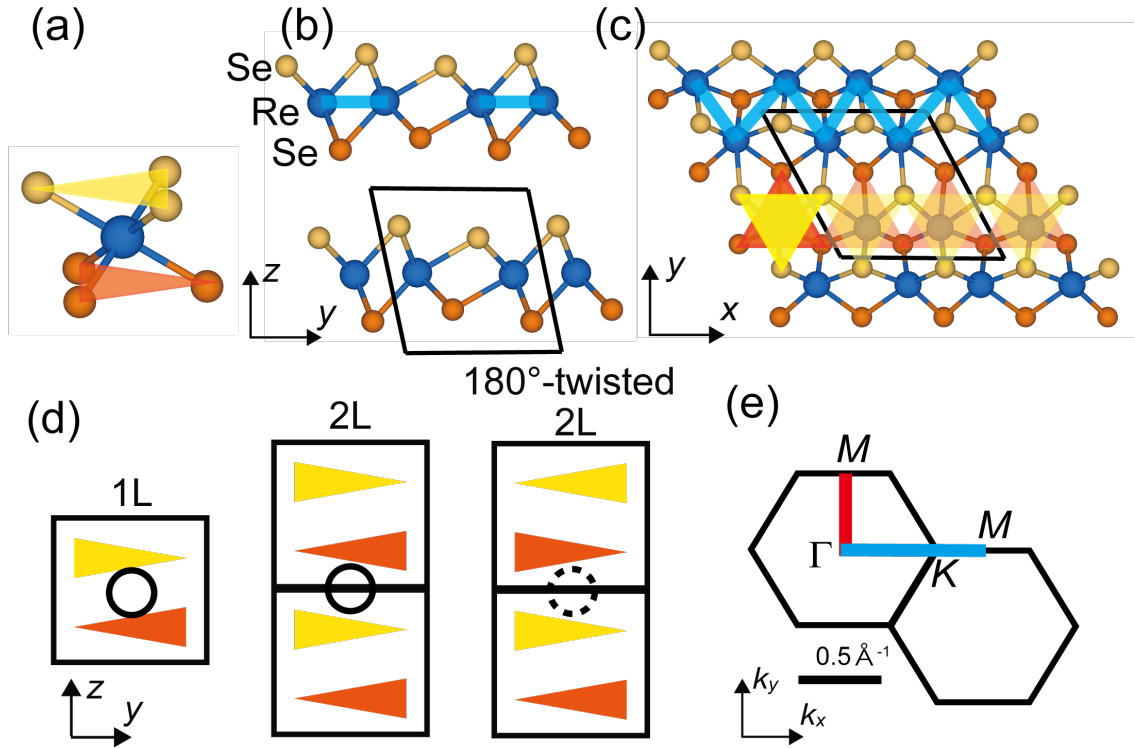

FIG. 1. (a) The typical non-distorted octahedral coordination for the T-type TMD. The yellow (orange) triangle indicates the orientation of the top (bottom) Se triangular networks. (b), (c) Side view (b) and top view (c) of the distorted T-type crystal structure of $ReSe_2$. The black frames indicate unit cells. The light blue segments represent the Re-zigzag-chain-like structure. (d) Schematic views of the crystal symmetries for monolayer (1L), bilayer (2L), and 180°-twisted 2L $ReSe_2$. The black rectangle represents each unit of the $ReSe_2$ layer. The black circles represent the inversion centers. The dotted circle depicted for the anti-parallel 2L $ReSe_2$ represents a lack of the inversion center by their stacking order. (e) Two-dimensional Brillouin zone for the 1L, 2L, and 180°-twisted 2L $ReSe_2$. The Brillouin zone is slightly shear-deformed from a regular hexagon. However, in this study, we represent the $\Gamma$-$K$ ($\Gamma$-$M$) as the direction parallel (perpendicular) to the Re zigzag chain because the actual high symmetrical $K$-point ($M$-point) locates only 0.008 (0.013) Å$^{-1}$ off the $k_x$ ($k_y$) axis.

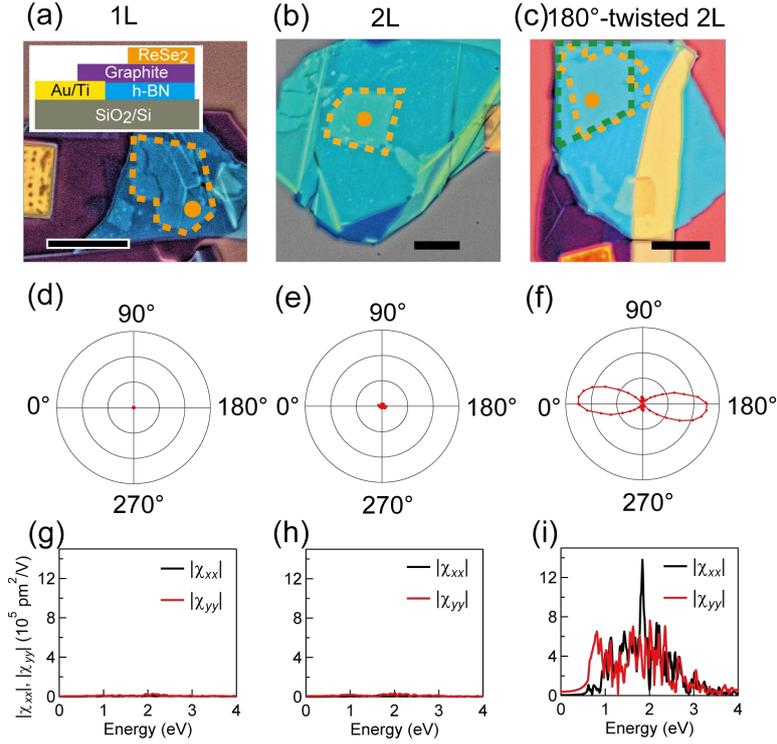

FIG. 2. (a)–(c) Optical microscope images of the samples for ARPES and SHG measurement. The inset in (a) shows a schematic of the sample. The black segments represent 10 μm. The monolayer ReSe$_2$ flakes used are outlined with orange and green broken lines. Orange circles indicate the SHG measurement position. (d)–(f) Polar plot of the SH intensity from 1L, 2L, and 180°-twisted ReSe$_2$. Linear-polarized component of the SHG parallel to the linear polarization of the incident light is detected. The measurement angles correspond to the optical microscope images in (a)–(c), respectively. For (f), the data between 0° and 180° are symmetrized and displayed in the region between 180° and 360°. (g)–(i) The calculated second order electrical susceptibility $\chi_{xx}$ and $\chi_{yy}$ for 1L, 2L, and 180°-twisted ReSe$_2$.

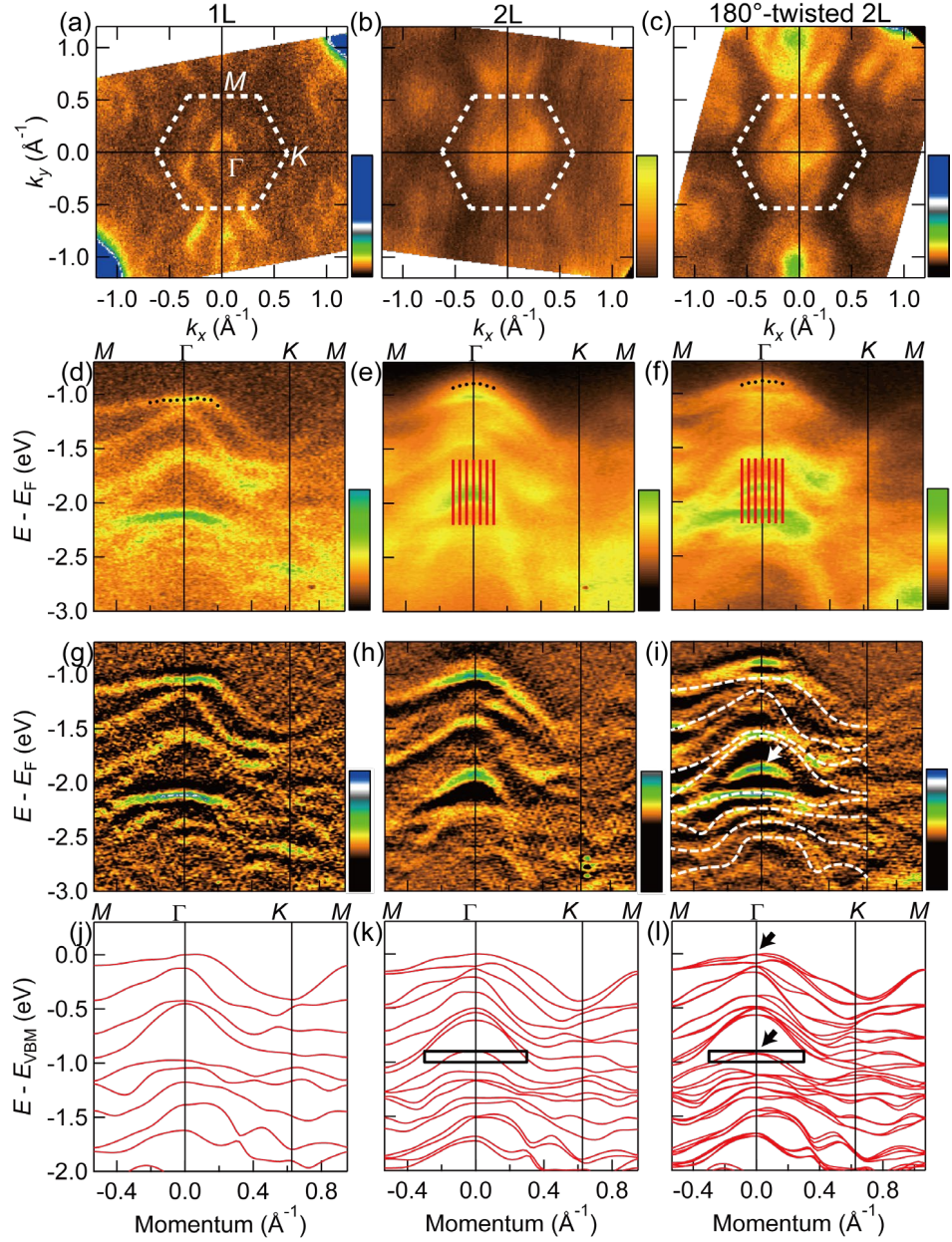

FIG. 3. (a)–(c) ARPES mapping images at a constant energy of $E - E_F = -1.3$ eV of 1L, 2L, and 180°-twisted 2L ReSe$_2$. White broken lines represent the Brillouin zones for 1L, 2L, and 180°-twisted 2L. (d)–(f) Combined ARPES images along the Γ-$M$ ($k_y$) and Γ-$K$-$M$ ($k_x$) directions. Those ARPES images along the $k_x$ ($k_y$) direction are symmetrized with respect to $k_x$ ($k_y$) = 0. Black circle markers represent the peak positions of the ARPES spectra for the respective highest valence bands. Red segments in (e) and (f) represent energy cuts for the energy distribution curves shown in Figs. 4(c) and 4(d), respectively. (g)–(i) Images obtained by the curvature analysis [46] for the ARPES images in (d)–(f). White broken lines in (g) represent the band dispersions of monolayer ReSe$_2$ obtained from the ARPES image, which are inevitably observed from the 180°-twisted 2L ReSe$_2$ samples. (j)–(l) Calculated band dispersions of 1L, 2L and 180°-twisted 2L ReSe$_2$. The origins of energy axes are set as the maximum of the valence band (VBM). The black rectangles in (k) and (l) indicate the area of the magnified view in Figs. 4(a) and 4(b), respectively.

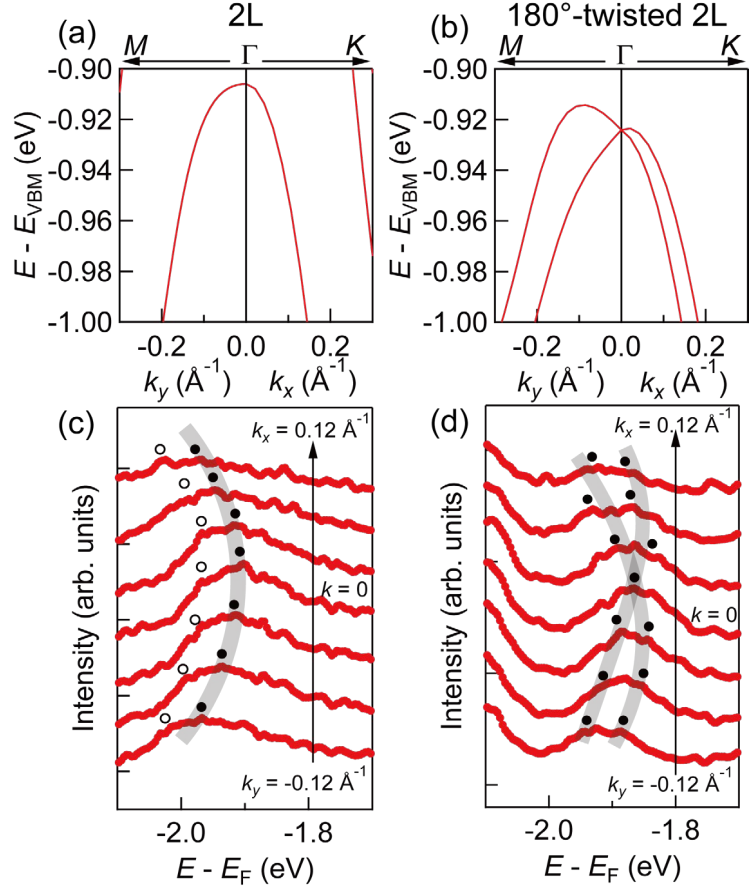

FIG. 4. (a), (b) Enlarged views of the calculational band dispersion in 2L and 180°-twisted 2L ReSe$_2$. The corresponding regions are indicated by the black rectangle in Figs. 3(k) and 3(l). (c), (d) EDCs extracted from the ARPES images of 2L and 180°-twisted 2L ReSe$_2$ as denoted by the red segments in Figs. 3(e) and 3(f), respectively. Black solid circle markers represent the peak positions, indicating signatures of possible spin-split band dispersions in 180°-twisted 2L ReSe$_2$. Open circle markers in (c) represent peak positions from area with different chemical potentials possibly.